\documentclass[10pt]{article}

\usepackage{amsmath}
\usepackage{amssymb}

\usepackage{graphicx}

\usepackage{cite}

\usepackage{color} 

\usepackage[labelfont=bf,labelsep=period,justification=raggedright]{caption}

\makeatletter
\renewcommand{\@biblabel}[1]{\quad#1.}
\makeatother

\date{}

\usepackage{url}  

\usepackage{tabularx}

\usepackage{framed}
\usepackage{nameref}
\usepackage{color}

\newenvironment{recommendation}
{
\begin{minipage}{\textwidth}
\quad
\begin{framed}
\noindent
  \textbf{\textsf{We recommend:\\}}%
}
{
\end{framed}
\end{minipage}
}%

\usepackage{setspace}
\usepackage{lineno}

\begin{document}

\begin{flushleft}
{\Large
\textbf{Privacy in Sensor-Driven Human Data Collection: \\ \vspace{0.05in}
\textit{A Guide for Practitioners}} \\ \vspace{0.09in}
\textcolor{red}{\textbf{Working Paper}}
}
\\ \vspace{0.1in}
Arkadiusz Stopczynski$^{1,2,\ast}$,
Riccardo Pietri$^{1}$,
Alex `Sandy' Pentland$^{2}$,
David Lazer$^{3}$, 
Sune Lehmann$^{1,4}$
\\ \vspace{0.1in}
\bf{1} \textbf{Technical University of Denmark}\\
\bf{2} \textbf{MIT Media Lab}\\
\bf{3} \textbf{Northeastern University}\\
\bf{4} \textbf{The Niels Bohr Institute}
\\ \vspace{0.1in}
arks@dtu.dk, riccardo.pietri@gmail.com, sandy@media.mit.edu, d.lazer@neu.edu, sljo@dtu.dk

\end{flushleft}

\tableofcontents
\newpage

\section*{Abstract}
In recent years, the amount of information collected about human beings has increased dramatically. 
This development has been partially driven by individuals posting and storing data about themselves and friends using online social networks (such as Facebook or Twitter) or collecting their data for self-tracking purposes (quantified-self movement). 
Data regarding human behavior is also collected through the environment, embedded RFIDs, cameras, traffic monitoring, business transactions, etc.. 
Across the sciences, researchers conduct studies collecting data with an unprecedented resolution and scale. 
Using computational power combined with mathematical models, such rich datasets can be mined to infer underlying patterns, thereby providing insights into human nature.

Much of the data collected is sensitive.
It is private in the sense that most individuals would feel uncomfortable sharing their collected personal data publicly.
For this reason, the need for solutions to ensure the privacy of the individuals generating data has grown alongside the data collection efforts.
Out of all the massive data collection efforts, this paper focuses on efforts directly instrumenting human behavior, and notes that---in many cases---the privacy of participants is not sufficiently addressed.
For example, study purposes are often not explicit, informed consent is ill-defined, and security and sharing protocols are only partially disclosed.

This paper provides a survey of the work related to addressing privacy issues in research studies that collect detailed sensor data on human behavior. 
Reflections on the key problems and recommendations for future work are included. 
We hope the overview of the privacy-related practices in massive data collection studies can be used as a frame of reference for practitioners in the field.
Although focused on data collection in an academic context, we believe that many of the challenges and solutions we identify are also relevant and useful for other domains where massive data collection takes place, including businesses and governments. 




\section{Introduction}\label{sec:introduction}
Humanity is recording the minute-to-minute details of human behavior and interaction at an unprecedented rate.
The datasets come from diverse sources such as user-generated content in online social networks (e.g.~Twitter, Facebook) and on other online services (e.g.~Flickr, Blogger); human communication patterns recorded by telecom operators and email providers; customer information collected by traditional companies and online wholesalers (e.g.~Amazon, Target); data from pervasive environments such as sensor-embedded infrastructures (e.g.~smart houses); social science experiments; the list continues.

As technology advances, the technical limitations related to storing and sharing these collections of information are gradually overcome, providing the opportunity to collect and analyze these digital traces.
This ever-increasing volume of user-generated content carries significant economic and social value and, as such, is of interest for business organizations, governmental institutions, and social science researchers.

Massive data collection efforts happen in virtually every aspect of the modern world; businesses, governments, and research institutions collect and mine data.
In many cases, this may not raise directly actionable privacy concerns.
Data may be publicly available, for example in the form of Twitter posts.
Or it may be owned and tightly controlled by a company, such as Call Detail Records (CDRs) collected by telecom operators.
Often, however, when data are collected by directly instrumenting human behavior, there is a need to ensure the privacy of the participants.

For the research community, the data revolution has had an impact that can hardly be underestimated. 
Data persistence and searchability combined with enhanced computational power has given rise to Computational Social Science (CSS), a domain of interdisciplinary research that gathers and mines large-scale behavioral datasets to study human behavior and social interactions~\cite{lazer2009life,eagle2009inferring,chronis2009socialcircuits}. 
In this paper we focus on the sensor-driven massive data collection efforts with human subjects for research purposes, as practiced in computational social science. 
Many of the challenges and solutions stated here translate, however, to other domains, such as classical social science or data collection in a business or governmental context.

Protecting the wellbeing of the participants of the studies in any domain is of utmost importance. 
The trust that exists between the participants and the researchers is even more difficult to establish and maintain when the amount and resolution of the collected data increases.
The more sensitive the data is, the more disastrous the consequences of misuse can be.
Protecting the privacy of the participants is driven by the ethics and responsibility of individual researchers; the risks are to the collective scientific endeavor, there is an issue of public confidence in the conducted research.
As we rapidly collect more and more data, even a single case of data abuse could lead to a loss of public confidence and --- as a consequence --- a decreased ability to carry out studies.
This is why we recommend a set of standards and tools for responsible management of privacy that can be easily deployed and used as more researchers join the big data movement.

Protecting human subjects is not a new idea.
Different research domains, countries, and institutions have their own means of ensuring proper treatment of human subjects.
In the United States, studies involving human subjects are regulated by the Institutional Review Boards (IRBs).
These academic committees are designated to approve, monitor, and review any medical and non-medical research involving humans. 
Part of the review process is an assessment of the informed consent procedure, ensuring that the participants are properly informed about the study.
The IRBs also provide guidelines for data confidentiality, such as removing identifying information, ensuring data deletion, or secure storage of the data.

However, as the collected data sets grow larger, more complex, and of higher resolution, the typical approach to treatment of human subjects may be insufficient.
Proper handling of massive data, informing and obtaining users' authorizations, ensuring privacy when performing data analyses; all these problems, although not new, become qualitatively more difficult to handle when the data become big.
We are seeing the dawn of studies that are regulated by the IRBs and corresponding bodies in other countries, but use methods which the IRBs are not proficient in assessing.
For example, in many of the research studies collecting massive amounts of data from human subjects, the data are accessed and analyzed in almost-real time, making it difficult to apply de-identification techniques that operate on the static datasets. 
Similarly, simply conveying the message to the participants about the study goals and scope can become a challenge, as the contact between the researchers and users decreases due to the scale of studies or the enrollment procedure, while the complexity of the data and subsequent analysis increases.
This poses a potential  problem for the research community wanting to do massive data collection on human subjects---studies that require IRB approval---when the IRB guidelines may not necessarily be suitable for such studies.

While classical social science studies primarily use questionnaires and interviews for data collection, many CSS studies use sociometetric badges~\cite{wu2008mining} or smartphones~\cite{pentland2008honest,raento2009smartphones,olguin2011mobile} to sense, collect, and transmit large quantities of multi-purpose data. 
Data collected from sensors include WiFi and Bluetooth IDs of the surrounding devices, GPS traces, SMS and call logs.
Data are also collected from online services (Facebook, Twitter), both from public streams and from users' accounts, based on their authorizations.  
While such a longitudinal approach allows scientists to maintain a broad scope, the scale and accuracy of the collected data often results in large amounts of sensitive information about the users, which may lead to privacy concerns.
Additionally, the sensitivity of the sensor data may be latent; it is not clear to the users why their location or social data need to be protected.  
We note that data of similar sensitivity has been historically collected in the health and biomedical fields, with collection of blood samples, brain scans, or genetic material.
With the development of online services such as \emph{23andMe} (\url{https://www.23andme.com/}) or \emph{patientslikeme} (\url{http://www.patientslikeme.com/}) and large available databases of medical data~\cite{gordon2005integrative}, the medical data privacy concerns are very similar to the sensor-driven human data on which we are focusing this article.

To a large degree, the public is unaware of the potential problems associated with sharing sensitive data. 
This lack of awareness is revealed in a number of contexts, for example via a documented tendency to `trade privacy for services' \cite{valueOfPrivacy}, or displaying carelessness regarding possible risks \cite{krumm2009survey,korth2011interdisciplinary}. 
It has been shown that while many users are comfortable with sharing daily habits and movements through multiple applications, only a minority of them are aware of which parties are the recipients of this information. 
Concurrently, pinpointing sensitive information about others is becoming easier using powerful search engines such as Google, Facebook Graph Search, or smartphone mashup apps (e.g.~Girls Around Me, \url{http://girlsaround.me/}).

To further aggravate this scenario, it has been shown that many of the techniques employed so far to protect users' anonymity are flawed. 
Scandals such as re-identification of users in the \textit{Netflix Prize} data set \cite{narayanan2008robust} and similar breaches \cite{sweeney2000simple,barbaro2006face} show that simple de-identification methods can be reversed to reveal the identity of individuals in those data sets \cite{de2013unique}. 
Attackers can also use these re-identification techniques to perpetrate so-called `reality theft' attacks  \cite{altshuler2011stealing}. 

In order to avoid such a negative scenario and to maintain and increase the trust relation between research community and participants, the scientific community must reconcile the benefits of their research with respect for users' privacy and rights. 
The current situation in the field is a heterogeneous set of approaches that raise significant concerns: study purposes are often not made explicit, the implementation of informed consent is often problematic, and in many cases, security and sharing protocols are only partially disclosed. 

As the number of participants and the duration of the studies increase, the pre-existing relation between researchers and participants grows weaker. 
Today, the participants in the largest deployments of the studies with human subjects are still students from a particular university, members of a community, or friends and family of the researchers. 
Studies that are open to the public allow for participants with no prior relation to the researchers. 
While this change certainly ameliorates some privacy concerns, it also lessens the personal investment in the wellbeing of study participants.
As a consequence, ensuring good practices in informing those participants about their rights, the consent they express, the incentives etc. becomes even more important. 

In a modern context, privacy implies that individuals should be able to determine how much personal information can be disclosed, to whom, and how it should be maintained and disseminated~\cite{warren1890right, westin1970privacy}.
Concurrently, we acknowledge that what is considered to be personal information -- and what is not -- changes among cultures, individuals, time, place, audience, and circumstances. 
Thus, the notion of privacy is dynamic, depends on context, and may change, for example as people discover that disclosing their private information may lead to value in return \cite{cate1997privacy}.

Providing adequate privacy for the participants in massive data collections studies, especially in a research context, is important.
We recognize that there are additional factors to consider in the design of studies, such as reproducibility, cost, reliability, and impact on the third parties.
It is important to find the right balance between all these aspects, which may not always be easy.
Although it may seem paradoxical, we see better privacy measures as a means for allowing more data to be collected, analyzed, and shared.
Users who feel confident about providing access to their data and see more value in doing so, will result in availability of even larger and more comprehensive datasets.   

Our contributions in this paper are two-fold.
First, we provide an overview of the privacy-related practices in existing sensor-driven human data collection studies.
We have selected representative works in the field and analyzed the fundamental privacy features of each one.
We focus on studies involving instrumentation of humans, where the data collection is sensor-driven and performed on a massive scale.
We include the consent-based collection of data from online social networks (OSNs), and additional sources, such as questionnaires, where such data augment the data collected from sensors.
Such collection efforts raise distinct issues around security, consent, and sharing.
The result is a longitudinal survey that we intend as a frame of reference for current and future practitioners in the field. 
Second, we lay the groundwork for a \textit{privacy management change process}. 
Using the review as a starting point, we have constructed a list of the most important challenges to overcome for the studies: we call these \textit{privacy actionable items}. 
For each item, we delineate realistic implementations and reasonable life-spans. 
Our goal is to inspire introspection and discussion, as well as to provide a list of concrete items that can be implemented today and that overcome some of the problems related to the current privacy situation.
The intended audience are researchers and scholars engaging in sensor-driven human data collection. 
We intend many of the reviewed practices and suggestions provided to generalize to other domains, such as data collection in companies or other types of data. 

\section{Sensor-Driven Human Data Collection in Academia}\label{sec:literatureReview}
We broadly categorize the projects selected for our privacy survey into two families: generic frameworks and specialized applications. 
The former category contains systems that collect a variety of different data streams deployed for the purposes of studying human behavior in longitudinal studies. 
The second category consists of applications that collect sensor data for a specific, well-defined purpose.

\paragraph*{Human Dynamics Group}
\textit{Friends and Family} is a data collection study deployed by the Massachusetts Institute of Technology (MIT) in 2011 to perform controlled social experiments in a graduate family community~\cite{aharony2011social}. 
For the purpose of this study, researchers collected 25 types of signals (e.g.~wireless network names, proximity to Bluetooth devices, statistics on applications, call and SMS logs) using Android smartphones as sensors.
\textit{Funf} -- the mobile sensing framework developed for this study -- is the result of almost a decade of studies in Media Lab Human Dynamics Group on data-collection using mobile devices. 
In 2008 a Windows Mobile OS~\cite{madan2011sensing} app was used to collect data from students and study connections between behavior and health conditions (e.g. obesity), and to measure the spread of opinions. 
Four years before that, a team from Media Lab studied social patterns and relationships in users' daily activities, using Nokia phones~\cite{eagle2006reality}. 
In 2003 a Media Lab team pioneered the field by developing the first sensing platform~\cite{eagle2003social} in order to establish how face-to-face interactions in working environments influence the efficiency of organizations. 
While purposes of the studies and mobile sensing technologies have evolved, the general setup with a single server to collect and store the data coming from the devices, has generally remained unchanged.

\paragraph*{OtaSizzle}
Another large-scale data-collection effort was conducted by Aalto University researchers in 2011~\cite{karikoski2011measuring}. 
Here, the focus was on understanding social relations by combining multiple data sources. 
The results showed that in order to better describe social structure, different communication channels should be considered.
Twenty students at the university were recruited by email invitation and participated in the experiment for at least three months. 
The research platform involved three different data sources: text messages, phone calls (both gathered with Nokia N97 smartphones), and data from an experimental OSN called \textit{OtaSizzle}, hosting several social media applications for Aalto University. 
All the gathered information was temporarily stored on the smartphones, before uploading to a central server.  

\paragraph*{Lausanne}
The last framework that we include here is the \textit{Lausanne Data Collection Campaign} (\textit{LDCC})~\cite{aad2010nrc,kiukkonen2010towards,laurila2012mobile}, conducted by the Nokia Research Center in collaboration with the EPFL institute of technology between 2009 and 2011. 
The purpose was to study users' socio-geographical behavior in a region close to Lake Geneva. 
The LDCC platform involved a proxy server used to collect raw information from phones, anonymizing the data before transferring them to a second server for research purposes.

\subsection{Smaller frameworks}
Below we present an overview of three groups of specialized platforms and smartphone applications developed by research groups for different purposes. 
In Table~\ref{tab:surveyedWork_generic} we present distributed architecture frameworks. 
Table~\ref{tab:surveyedWork_policy} shows projects related to the creation of privacy policies and management. 
In Table~\ref{tab:surveyedWork_specific} we present applications that generate, collect, and share information about users, using smartphones as sensing devices. 
Other frameworks are also cited to provide useful examples. 
We remark that it is not our interest to discuss the primary goals of the mentioned studies (incentives, data mining algorithms, or results), but to present an overview of architectures, procedures and techniques employed for data collection and treatment---with specific focus on privacy.

\begin{center}
\begin{table}[h!]
\footnotesize
{\renewcommand{\arraystretch}{1.2}
\begin{tabularx}{\linewidth}{|p{3cm}|X|X|}

\hline 
Name  & Purpose & Privacy measures\\ 
\hline 

\textbf{Vis-a'-Vis} \cite{shakimov2011vis} 2011 - Duke University, AT\&T Labs &
A personal virtual host running in a cloud computing infrastructure and containing users' (location) information. &
Allows users to manage their information directly from their virtual host with full control; exposes \textit{unencrypted} data to the storage providers. \\
\hline 

\textbf{Confab} \cite{hong2004architecture}  
2004 - University of California at Berkeley, University of Washington & 
A distributed framework facilitating development of other privacy-aware applications for ubiquitous computing. & 
Personal data is stored in computers owned by the users, providing greater control of information disclosure. \\ 
\hline 

\textbf{MyLifeBits} \cite{gemmell2006mylifebits,gemmell2002mylifebits} 
2001 - Microsoft Research &
Early example of digital database for individuals' everyday life, recording and managing a massive amount of information such as digital media, phone calls, meetings, contacts, health data etc. &
Information kept in SQL databases. Privacy concerns mentioned but \textit{not} addressed in the project.\\
\hline 

\textbf{VPriv} \cite{popa2009vpriv} 
2009 - MIT, Stanford University & 
Privacy-aware location framework for car drivers, producing an anonymized location database. Can be used to create applications such as usage-based tolls, automated speeding tickets, and pay-as-you-go insurance policies. &
\textit{Homomorphic encryption} \cite{rivest1978data}, ensures that drivers' identities are never disclosed in the application.
\\
\hline 

\textbf{HICCUPS} \cite{molina2009hiccups} 
2009 - University of Massachusetts Amherst & 
A distributed medical system where a) physicians and caregivers access patients' medical data; b) researchers can access medical aggregate statistics & 
Implements homomorphic encryption techniques to safeguard patients' privacy.\\
\hline 

\textbf{Darwin} \cite{miluzzo2010darwin} 
2010 - Dartmouth College, Nokia &
A collaborative framework for developing a variety of sensing applications, such as place discovery or tagging applications. & 
Provides distributed machine learning algorithms running directly on smartphones. Raw data are not stored on and do not leave the smartphone. \\
\hline 

\textbf{AnonySense} \cite{cornelius2008anonysense,kapadia2008anonysense}  
2008 - Dartmouth College & 
An opportunistic framework for applications using multiple smartphones to accomplish a single sensing task. & 
Provides anonymity to the users deploying $k$-anonymity \cite{sweeney2002k}.\\
\hline 

\end{tabularx}
}
\caption{\textit{Distributed frameworks}. The \textit{Vis-a’-Vis}, \textit{Confab}, and \textit{MyLifeBits} frameworks are personal information collectors that play the roles of users' virtual aliases. Two implementations of homomorphic encryption for drivers and healthcare follow. Finally, \textit{Darwin} and \textit{AnonySense} are collaborative frameworks.}
\label{tab:surveyedWork_generic}
\end{table}
\end{center}

\begin{center}
\begin{table}[h!]
\footnotesize
{\renewcommand{\arraystretch}{1.2}
\begin{tabularx}{\linewidth}{|p{3cm}|X|X|}

\hline 
Name  & Purpose & Privacy measures\\ 
\hline 

\textbf{PViz} \cite{mazzia2012pviz} 
2012 - University of Michigan &
A graphical interface that helps users of social networks with policy comprehension and privacy settings. &
Nodes represent individuals and groups, different colors indicate the respective visibility. \\
\hline 

\textbf{Virtual Walls} \cite{kapadia2007virtual} 
2007 - Dartmouth College, University of St Andrews &
A policy language that leverages the abstraction of physical walls for building privacy settings. & 
Three levels of granularity ("wall transparencies") allow users to control quality and quantity of information disclosure
towards other digital entities (users, software, services). \\
\hline 

\textbf{A policy-based approach to security for the semantic web} \cite{kagal2003policy} 
2003 - University of Maryland Baltimore Country &
A distributed alternative to traditional authentication and access control schemes.&
Entities (users or web services) can specify their own privacy policies with rules associating credentials with granted rights (access, read, write, etc.). \\
\hline 
 
\end{tabularx}
}
\caption{\textit{Policy frameworks}. An overview of tools that help users to understand and control their policy settings.}
\label{tab:surveyedWork_policy}
\end{table}
\end{center}

\begin{center}
\begin{table}[h!]
\footnotesize
{\renewcommand{\arraystretch}{1.2}
\begin{tabularx}{\linewidth}{|p{3cm}|X|X|}

\hline 
Name  & Purpose & Privacy measures\\ 
\hline 

\textbf{CenceMe} \cite{miluzzo2008sensing} 
2008 - Dartmouth College &
Uses smartphones to sense peoples' activities (such as dancing, running, ...) and results are automatically shared on Facebook.&
As soon the classification is performed on the devices the data is erased.\\
\hline 

\textbf{GreenGPS} \cite{ganti2010greengps} 
2010 - University of Illinois &
A GPS navigation service which discovers \textit{greener} (fuel-efficient) paths through drivers participatory collaboration (based on the previous framework \textit{Poolview} \cite{ganti2008poolview}). & 
No fine-grained data control: if users feel the need for privacy, they need to switch off the GPS device to stop data collection. \\
\hline 

\textbf{Speechome Recorder} \cite{vosoughi2012portable, roy2006human} 
2012 - MIT, Northeastern University &
An audio/video recording device for studying children's daily behavior in their family house.  &
Ultra-dense recordings temporarily kept locally and uploaded to central server. Scarce information about data encryption and transport security protocols are provided. \\
\hline 

\textbf{Cityware} \cite{kostakos2008cityware} 
2008 - University of Bath &
Application for comparing Facebook social graph against real-world mobility traces detected using Bluetooth technology. &
Switching Bluetooth to invisible as a way to protect users' privacy.\\
\hline 

\textbf{FriendSensing} \cite{quercia2009friendsensing} 
2009 - MIT, University College London &
Bluetooth used to suggest new friendships evaluating device proximities. &
Same as \textit{Cityware}.\\
\hline 

\textbf{FollowMe} \cite{ypodimatopoulos2010follow} 
2010 - Massachusetts Institute of Technology & 
Service that uses HTTP and Bluetooth to automatically share indoor position (malls, hospitals, airports, campuses). &
Implements a decentralized architecture to improve users' location privacy.\\
\hline 

\textbf{Locaccino} \cite{toch2010empirical} 
2010 - Carnegie Mellon University &
A mobile sharing system created to study peoples' location privacy preferences. &
Relevant privacy considerations will be reported later in the article.\\
\hline 

\textbf{Bluemusic} \cite{mahato2008implicit} 
2008 - RWTH Aachen, University of Duisburg-Essen &
Application developed for studying personalization of public environments. It uses Bluetooth \textit{public} usernames as pointers to web resources that store users preferences.& 
Same as \textit{Locaccino}.\\
\hline 

\end{tabularx}
}
\caption{\textit{Applications.} Although providing great functionality for the users, the privacy-oriented settings for the user are often not sufficiently implemented.}
\label{tab:surveyedWork_specific}
\end{table}
\end{center}


\section{Informed Consent}\label{sec:permissionAndConsent}
Here we examine how the state-of-the-art massive data collection studies involving human subjects approach the participants' understanding of the consequences of data collection, as well as their ability to control and access personal data during and after such studies.

In academic studies, informed consent consists of an agreement between researchers and the data producer (user, participant) by which the latter confirms she understands and agrees to the procedures applied to her data, including collection, transmission, storing, sharing, and analysis. 
The intention of the informed consent is to protect individual autonomy, ensure voluntary participation, and for the users to comprehend which information will be collected, who will have access to that information, what the incentive is, and for what purposes the data will be used~\cite{friedman2005informed, beauchamp2001principles}. 
In all studies the research ethic has a central role in relation to protecting the privacy of participants. 
For this reason, to be able to perform analyses on the collected data, scientists typically work under agreements defining allowed uses~\cite{laurila2012mobile,karikoski2011measuring}.

The informed consent procedure was introduced in the Nuremberg Code following the revelations of torture by Nazi doctors during World War II~\cite{utley1992nazi}.
Originally conceived for medical studies, informed consent has since been studied from the moral, ethical, philosophical, psychological, or practical point of view~\cite{faden1986history, applebaum1987informed}, understood as a cornerstone of the rights of patients and subjects. 
Informed consent imposes legal obligations on the researchers, covering researcher-participant or doctor-patient relationship, use of consent forms, refusal of treatment, or participation in the research~\cite{berg2001informed}. 
Protecting personal privacy is meant to protect individuals against stigmatization and discrimination, as well as various forms of social control which may undermine personal autonomy and democracy itself~\cite{faden1986history,beauchamp2001principles}. 
To be specific, there are many ways in which abuses might be directly targeted---from imposing higher insurance rates based on individual shopping history \cite{gittleson2013big} or creating problems for society as a whole, by limiting users' choices and enclosing them in information bubbles~\cite{hannak2013measuring}.

In the sensor-driven human data collection studies, we note the scarcity of published informed consent procedures; the majority of the studies we reviewed have not published their consent procedures~\cite{madan2011sensing,eagle2009inferring,ypodimatopoulos2010follow,madan2010social,chronis2009socialcircuits,miluzzo2008sensing,aharony2011social,madan2011pervasive,olguin2009sensible}. 
Due to this, it is difficult to produce comparisons and to create templates applicable for future studies. 
Where the procedures for achieving informed consent are reported, the agreement was carried out using forms containing users' rights (similar to \url{http://green-way.cs.illinois.edu/GreenGPS_files/ConsentForm.pdf}, e.g~\cite{kiukkonen2010towards,ganti2010greengps,kotz2009privacy,eagle2006reality,madan2011sensing,karikoski2011measuring}) or by accepting the terms of use during the installation of an application.

\subsection{Understanding}
Presenting extensive information about a study does not guarantee that informed consent has been implemented successfully: years of End-User License Agreements (EULAs) and other lengthy legal agreements show that most individuals tend to blindly accept forms that appear before them and to unconditionally trust the validity of the default settings which are perceived as authoritative~\cite{bohme2010trained,ploug2013agreeing,ploug2013informed}.
The problem of understanding the consent procedure has been extensively studied in the medical context.
Cassileth et al. in 1980 explored how much patients understand and are able to recall the information presented on the consent forms and in oral explanation, finding that only 60 percent of the participants understood the purpose and nature of the procedure~\cite{cassileth1980informed}.
The patients in the study believed that the consent forms were meant to ``protect the physician's rights'' and authors concluded that, even with comprehensible forms, the legal connotations led to cursory reading and inadequate recall. 
A study of readability of the informed consent forms in 2003 found that on the Flesch–Kincaid scale (0 - 12 US grade level) the average readability score for text provided by the IRBs was 10.6, with 95\% falling into the 10.3 - 10.8 interval~\cite{paasche2003readability}.
At the same time, among the 61 schools that specified explicitly grade-level standard for their forms, only 8 percent met this standard; the average readability score exceeded the stated standard by 2.8 grade levels.
This indicates potential issues in how well the participants can comprehend the text they are signing, based on language comprehension alone, without even considering required domain-specific knowledge.
Sugerman et al. reviewed in~\cite{sugarman1999empirical} the informed consent procedures for older adults, concluding that older age and fewer years of education contributed to decreased understanding of the consent form.

Various means of improving the understanding of the consent and making it truly informed have been researched, primarily in the clinical context.
Lavori et al. explored in~\cite{lavori1999improving} the possibility and challenges of improving informed consent based on experimentation in ongoing clinical trials with respect to ethical, organizational, conceptual, and technical aspects.
Evaluating how people understand their privacy conditions can be done, for example, by conducting feedback sessions throughout the duration of the experiment~\cite{kiukkonen2010towards,mahato2008implicit}.
One improvement could be to allow users to gradually grant permission over time, but the efficacy of this approach is not clear: some studies have shown that users understand the issues about security and privacy more clearly when the requests are presented gradually~\cite{egelman2012choice}, while others argue that too many warnings distract users~\cite{kapadia2007virtual,felt2011effectiveness,cranor2006they}. 
Flory and Emanuel presented a comprehensive review of the methods that can be used to improve the understanding of the consent in~\cite{flory2004interventions}. 
They found that in the clinical context, participants may frequently not understand the disclosed information; one of the requirements of a well-implemented informed consent procedure~\cite{national1978belmont, saif2000world}. 
To improve the consent procedure, they point out that the multimedia augmentation or replacement of the consent form can significantly improve participants' understanding; the effect is most pronounced in the group of mentally ill patients~\cite{dunn2002improving, agre2003improving}, although not always manifesting in the general population~\cite{fureman1997evaluation, weston1997evaluating}.
These results suggest that in the examined (clinical) context, basic comprehension of the form was not the problem, but rather higher-level reasoning about the information presented.
Similarly, the effect of simplification of the consent forms is not clear, and depends to a large extent on whether the form is the only source of information for the participant.
The most effective measure to improve understanding seems to be person-to-person contact and discussion about the research scope, purpose, procedures etc.
Although this result is not surprising, in the context of this article the usefulness of such person-to-person interventions is limited; with hundreds and thousands of participants in the massive sensor-driven data collection, explaining the study to each one of them may simply not be feasible.

\subsection{Control}
In most cases, the current informed consent techniques represent an all-or-nothing approach that does not allow the user to select subsets of the permissions, making it only possible to either participate in the study fully or not at all~\cite{felt2011effectiveness}.
Accounting for cases where the user are only to contribute subsets of the data is usually done by allowing to skip questions in the questionnaires or decline to participate in the optional elements of the study~\cite{aharony2011social}.
Generally, once the consent is granted by the user, her data contribution to the dataset becomes the property of the researchers, in that they can analyze, modify, or redistribute the data as they see fit, depending on the terms of the consent, but typically simply provided that basic de-identification is performed. 
As we suggest in the section called~\nameref{sec:informationOwnership}, it is a good practice for the researchers to clarify the sharing schemes and expiration of the collected information to the participants: if users cannot follow the flow of their data, it is difficult to claim that truly informed consent has been expressed. 
Since so little is understood about the precise nature of conclusions that may be drawn from high-resolution data, it is important to continuously work to improve and manage the informed consent as new conclusions from the data can be drawn. 
One option is that the paradigm should evolve from a one-time static agreement to dynamic consent management~\cite{shabajee2006informed}. 
Furthermore, the concerns related to privacy are context-specific~\cite{toch2010empirical,lindqvist2011m} and vary across different cultures~\cite{avancha2009privacy,mahato2008implicit}. 
In the literature, the need for a way to let the users easily understand and specify which kinds of data they would like to share and under what conditions was foreseen in 2002 by the W3C group, with the aim to define a Platform for Privacy Preferences (P3P) (suspended in 2006), in 2003 by Kagal et al.~\cite{kagal2003policy}, and also in 2005 by Friedman et al.~\cite{friedman2005informed}, all shaping dynamic models for informed consent. 
Recent studies such as~\cite{toch2010generating} have worked to design machine learning algorithms that automatically infer policies based on user similarities. 
Such frameworks can be seen as a mixture of recommendation systems and collaborative policy tools in which default privacy settings are suggested to the user and then modified over time. 

\subsection{Living Informed Consent}

We propose the term \textit{Living Informed Consent} for the aligned business, legal, and technical solutions whereby the participant in the study is empowered to understand the type and quality of the data that is being collected about her, not only during the enrollment, but also as the data is being collected, analyzed, and shared with third parties. 
Rather than a pen and clipboard approach for the user enrollment, participants should expect to have a virtual space (a website or application) where they can change their authorizations, drop-out from the study, request data deletion, as well the ability to audit who is analyzing their data and how frequently their data are accessed. 

As the quantity and quality of the data collected increase, it becomes difficult to claim that single-sentence description of \textit{we will collect your location} truly allows the participant to realize the complexity of the signal collected and possible knowledge that can be extracted from it. 
For example, even a very small number of data points, such as location updates, can uniquely identify a participant, as has been shown in~\cite{de2013unique}. 
For this reason, access to the extracted high-level features of the data should be favored, as outlined in the openPDS approach~\cite{de2012trusted}.
High-level features are not only inherently better for privacy, as they limit the possible non-authorized uses of the data, they also make the authorization process easier for the user to comprehend (e.g. rather than agreeing to share location data, the user can agree to share the cities she visits, or average distance traveled per hour). 
Such an engaging approach to the users' consent will also be beneficial for the research community because as the relation with the user in terms of their consent expression extends beyond initial enrollment, the system proposed here makes it possible for the user to sign up for the new studies and donate their data from the previous studies.

The concept of Living Informed Consent is compatible with the discussion that has recently started around the need for improvements in the consent procedures; primarily in the biomedical research.
Laura Siminoff pointed out in~\cite{siminoff2003toward} a troubling lack of research and dialog around informed consent.
A growing number of practitioners adopt the belief that current procedures are but an illusion of giving control to the users.
In similar spirit, Erika Check Hayden discussed in~\cite{erika2012informed} how the informed consent procedures are today broken: not giving participants sufficient control over their data and participation and not including the future developments allowing us to extract more from the data. 
At the same time the bureaucracy around obtaining the consent is growing to be an obstacle for high-quality research.
Even the issue of sharing the results back to the participants is not straightforward; patients should have the option not to learn the results of research that was done on their data, such as genome.
The reuse of the data for different research is challenging; data from the participants either become part of large datasets maintained by researchers, where users effectively lose control over them, or data can only be used for particular analysis, specified in the consent.
Erika Hayden concludes that we can expect approaches allowing participants to control and track how researchers use their data. 
The same topic has been the focus of the Editorial in Nature vol. 486 entitled \emph{Time to open up}~\cite{time2012informed}.
Addressing the problem of broken consent procedure in biomedical studies, the article calls:

\begin{quote}
\emph{
Technology and some creative thinking should be able to provide a solution to this problem. In an era when people can control who sees what types of personal information on their Facebook page, programmers should be able to design similar tools to give research volunteers a degree of influence over who uses their data and for which types of research.
}
\end{quote}

Lunshof et al. propose in~\cite{lunshof2008genetic} an opposite approach to solve the same problem in the field of genomics.
Rather than pushing for more end-user control, their thesis is that true confidentiality is impossible, due to the inherent uniqueness of the data.
In their \emph{Open Consent} the authors propose that volunteers will consent to unrestricted re-disclosure of data and disclosure of the gained knowledge, even if it cannot be predicted what can be learned from the data at the time of consent.
The consent does not invoke promises of anonymity, privacy, or confidentiality, as those properties are impossible to truly guarantee for the dataset.
Instead, the use of the data and disclosure practices should be based on ethics and the principle of veracity, backed by the reputation of researchers and their institutions.
While the volunteers can request the deletion of their data, no guarantees are given about full removal, as the data might have contributed to published work, that cannot easily be reversed (see our discussion inthe ~\nameref{sec:reproducibility} section).
Open Consent is used in the \emph{Personal Genome Project}, in which the participants are considered informed volunteers, willing to give up their privacy for the public good, a practice causing some controversy~\cite{ian2013pgp}.

The Living Informed Consent approach we propose, is aimed for the data that is most valuable when accessed in real-time and continuously: mobility patterns, health state, social activity.
For such data we may not want to operate on frozen, aging datasets; instead we access  fresh data, both for research purposes and to build applications feeding information back to the participants, either as a behavior-change mechanism or to provide value for participation.
Rather than clinical practice, where it is feasible and clearly beneficial to engage in person-to-person explanation of the research process, in sensor-driven massive data collection the participants may be too separated from the researchers to make such procedures feasible.
At the same time, the expectations of the participants on sensor-driven studies are generally lower when compared to clinical research; the problem of participants' vulnerability and pressure to join the research is significantly smaller.
For these reasons, we see Living Informed Consent, backed up by technical solutions such as openPDS as a viable framework for addressing the rising problems of consent procedure.

\subsection{Living Informed Consent and Reproducibility}
\label{sec:reproducibility}

Giving the participants full control over their data can cause concerns about research replicability and reproducibility. 
If the data can change at any time, either because users modify or delete them, the results published on the basis of the data become impossible to replicate.
One possible solution is to `freeze' the data used for each publication, freezing precisely the data needed to replicate the published results.
The issue of replicability must be addressed by the Living Informed Consent procedure.
As with all the other data access, the freezes of the data---or preferably the relevant features needed to replicate the results---should be done subject to user authorizations, but once a freeze is done, the data must remain unchanged.
Thus, for example, users may still request the deletion of their newer data, but cannot request modifications or deletions from the frozen copy of the dataset.
Freezes should not be part of the main database and should not be used for active research, but archived to enable existing results to be revisited.

There is an ongoing discussion about the nature of desired replicability and reproducibility in various domains. 
Chris Drummond suggested in~\cite{drummond2009replicability} a distinction between these two terms, based on whether the focus is on repeating the experimental setup and subsequent findings (replicability) or the results themselves (reproducibility).
Cases of dishonesty, incompetence, or error do happen.
The prominent examples in recent years include a programming error in the calculation of the impact of debt on economic growth~\cite{reinhart2010growth} or insufficiently controlled statistics in a study examining the impact of abortion on crime rates~\cite{donohue2001impact}.
Similarly, the interpretation of the collected massive data may be problematic.
Bertrand et al. in~\cite{bertrand2013sentiment} examined the sentiment of New York City on the basis of collected Twitter messages.
In the following articles in the popular press~\cite{ny2013}, the authors identified the saddest spot of the city as Hunter College High School.
This has proven to be a mistake in the data interpretation, where tweets from a single account in close proximity to the school greatly biased the results.
Being able to check and identify such errors is only possible if the underlying data is available.
Although it is important to have access to the exact description of the experiment or a study, as well as the collected data or extracted features on which the algorithms can be applied to produce the same results (replicability), it may be argued that, beyond cases of scientific fraud or incompetence, such replication does little to strengthen the results.
The value of reproducibility typically arises from deriving the same results from different experimental setups different from each other. 
Mina Bissell points in~\cite{bissell2013reproducibility} to the challenges of replication in a biomedical context, due to the required cost, expertise, sensitivity to conditions etc.
We can see how the concern is valid in various domains, including massive data collection efforts, where the cost and time required to collect datasets, although diminishing, may prohibit the exact replication efforts.

We believe that frozen datasets from massive collection studies, allowing for replication to the last decimal place by using the same algorithms, can be a solution for the important issue of transparency of science and ensuring that re-visiting the findings is possible.
It is, however, even more significant to be able to reproduce the results in different settings, which by definition will vary, at least slightly, in population, setup, duration etc.
In this context, a small fraction of users dropping out of the study and deleting their information from the published dataset should not harm the replicability, as the results should be robust to such small perturbations.
There is currently no good answer to the problem of balancing the privacy of the users --- inherent right to decide about data access authorizations --- and the availability of the data for the purpose of reproduction of the results and general transparency.
The solution will likely require both legal and technical frameworks for making the data available for the purpose of reproduction, but it will require new authorizations for other purposes.

\begin{recommendation}
A simple yes/no informed consent option does not live up to the complex privacy implications related to studies of human behavior. 
For that reason, we believe that users should play a more active role in shaping their involvement in such studies.
This view gains support from studies showing that people do not, in general, realize smartphone sensing capabilities nor the consequences of privacy decisions~\cite{klasnja2009exploring,felt2012android,krumm2009survey}.

Additionally, we suggest to carefully consider special cases where the participants may not have the competence or the authority to fully understand the privacy aspects~\cite{shapiro2013regulation,underwood2012blackberry,vosoughi2012portable,stutzman2013silent}.
For these reasons we urge special caution in cases where data collection is performed for unspecified later use without clearly stated study purposes, as well as when information about what long-term plans for the data collection is not clearly laid out (e.g. what happens to the data once the study is concluded)~\cite{friedman2005informed}. 
In short, we propose a move towards adapting the principles of Living Informed Consent in studies involving human subjects.
\end{recommendation}

\section{Data Security}\label{sec:dataSec}
The security of the collected data, although necessary for ensuring privacy goals, is not often discussed when reporting on collection of large-scale sensor-driven human data~\cite{ganti2010greengps, miluzzo2010darwin, miluzzo2008sensing,kotz2009privacy,vosoughi2012portable,karikoski2011measuring,shakimov2011vis}.
In the next sections we illustrate how security has been addressed in the current, centralized frameworks and how it can be integrated in (future) distributed solutions; what follows is not an exhaustive list, but a compendium of important techniques that can be applied for frameworks for data collection human subjects, as well as potential attacks to consider.

\subsection{Common Solutions}\label{sec:centralServer}
The centralized architecture, where the data is collected in the form of a single dataset (typically stored in a single database), has been the preferred solution in the majority of the surveyed projects~\cite{
vosoughi2012portable,
underwood2012blackberry,
kapadia2007virtual,
popa2009vpriv,
madan2011pervasive,
madan2010social,
kiukkonen2010towards,
miluzzo2008sensing,
ganti2010greengps,
vosoughi2012portable,
roy2006human,
aharony2011social,
olguin2009sensible,
eagle2006reality,
madan2011sensing,
kostakos2008cityware,
quercia2009friendsensing,
toch2010empirical,
mahato2008implicit,
karikoski2011measuring}. 
When a data package reaches the central server, information is usually stored in SQL databases (e.g.~\cite{aharony2011social,
gemmell2006mylifebits, gemmell2002mylifebits,
madan2010social,
madan2011pervasive}) and aggregated for later analysis. 

Centralized architectures suffer from several security problems. Availability is difficult to guarantee when the server becomes overloaded with heavy traffic or subject to a Denial-of-Service attack and, if compromised, a single server can reveal all user data at once.

Malware and viruses for mobile phones are on the rise; given the amount of sensitive information present on the devices, there is a need for mobile applications that take this reality into consideration in order to avoid loss of personal data~\cite{altshuler2011stealing}. 
To address this problem, some of the frameworks reduce the time the raw sensor information is stored on the device, for example, the \textit{Darwin} platform discards data records once a classification task has been performed. 
Such approaches, however, are not the norm, most of the sensing applications we surveyed use an opportunistic approach to data uploading, pushing data only when the user is connected to WiFi. 
As a consequence, devices may end up with large amounts of data stored on the device memory~\cite{madan2011pervasive}, which may introduce a security threat if the devices do not use encrypted file-systems by default. 
A possible way to tackle this problem is by employing frameworks (such as FunF ~\cite{aharony2011social}) that provide reliable data storage systems by encrypting files before moving them to phone memory by minimizing the amount of data transmitted~\cite{miluzzo2008sensing}. 

Data transmission should be realized over HTTPS (HTTP Secure) to provide protection in the transport layer, such as for example in~\cite{shakimov2011vis,
cornelius2008anonysense, kapadia2008anonysense,
kotz2009privacy}. We note that some of the surveyed frameworks transmit data over plain HTTP~\cite{
hong2004architecture,
miluzzo2008sensing,
ganti2010greengps,
ypodimatopoulos2010follow,
mahato2008implicit}. Although this does not automatically indicate security problems, it is a good practice to use transport layer encryption and the protection it offers.

Even encrypted content can disclose information to malicious users, for example when the traffic flow is observed.
The opportunistic architecture of transmission and battery constraints does not allow smartphones to mask communication channels with dummy traffic to avoid analysis of the communication patterns (traffic analysis)~\cite{hong2004architecture,cornelius2008anonysense}.

\begin{recommendation}
We note that in all of the surveyed frameworks~\cite{shakimov2011vis, 
hong2004architecture,
gemmell2006mylifebits,
gemmell2002mylifebits,
molina2009hiccups,
miluzzo2010darwin,
cornelius2008anonysense,
kapadia2008anonysense,
kotz2009privacy,
kostakos2008cityware,
ypodimatopoulos2010follow}, the mechanisms for access control, user authentication, and data integrity checks (when present), were implemented specifically for the purpose of a given study. 
While `closed' custom solutions are the norm, they pose an intrinsic problem for privacy due to the complexity of the issue. 
No single group can be expected to solve every aspect of data security, hence both openness and collaboration are necessary, so that researchers do not have to create new security platforms for every new study. 

Finally, we recommend that the security of the entire platform (network and server) is enhanced by adhering to standard checklists, as illustrated in the guidelines on firewalls~\cite{scarfone2009guidelines} and intrusion detection systems~\cite{scarfone2007guide} by the National Institute of Standards and Technology (NIST).
\end{recommendation}

\subsection{Distributed Architecture}\label{sec:cloud}
In recent years, it has been a trend for businesses to store data in highly distributed architectures, or even off-site, in the `cloud'. 
We understand cloud as any remote system which provides a service relying on the use of shared resources~\cite{hay2011storm,fernando2012mobile}. 
An example can be a storage system which allows the users to backup their files and ubiquitously access them via the Internet (e.g. Dropbox) or a service for renting servers and related resources (e.g. Amazon Web Services).

Apart from facilitating the processes of data storage and manipulation, employing cloud solutions could improve the overall security of the collection efforts. 
As noted above, it is currently typical for platforms to be designed and implemented from scratch, in an environment where thorough testing with respect to security may not be a priority. 
If server platforms such as Amazon EC2 are used in the data collection frameworks, security mechanisms such as access control, encryption schemes, and authorization can be enforced in standard and well-tested ways. 
This primarily applies to moderately sensitive data that needs to be accessed in real-time, when keeping it on an offline tightly controlled machine is not feasible.

Using cloud solutions can make it easier to create data collection frameworks that allow users to \textit{own} their personal information. 
This allows for continuous monitoring of the personal data, controlling the access, and verifying deletion, which in turn may make users more willing to participate in the studies. 
One possible way to achieve such personal data stores is to upload the data from the mobile devices to personal datasets (e.g.~personal home computers, cloud-based virtual machines) as shown in \textit{Vis-a'-Vis}, \textit{Confab}, \textit{MyLifeBits} platforms, rather than to a centralized system. 
On the one hand, with these electronic aliases, users will feel more in control of their personal data, diminishing their concerns about systems that centralize data. 
On the other hand, part of the security of users' own data will inevitably rely on the users themselves -- and on the service providers who manage the data. 


Given the sensitive nature of the data, vulnerabilities in the cloud architectures can pose serious risks for the studies and, while cloud solutions might provide an increased level of security, they are definitely not immune to attacks (see~\cite{calluru2010exploitation} for attacks taxonomy and~\cite{hashizume2013analysis} for a general analysis on the cloud security issues). 

Sharing resources is a blessing and a curse of cloud computing; it helps to maximize the utility/profit of resources (CPU, memory, bandwidth, physical hardware, etc.), but at the same time it makes it more difficult to assure security, as both physical and virtual boundaries have to be reinforced.
The security of the Virtual Machines (VMs) becomes as important as the physical security because ``any flaw in either one may affect the other''~\cite{hashizume2013analysis}.
Since multiple virtual machines are hosted on the same physical server, attackers might try to steal information from one VM to another using cross-VM attacks~\cite{ristenpart2009hey}. 
One way to violate the data confidentiality is to compromise the software responsible for coordinating and monitoring different virtual machines (hypervisor) by replacing its functionalities with others aimed at breaching the isolation of any given pair of VMs, a so-called Virtual Machine Based Rootkit~\cite{king2006subvirt}.

Another subtle method to violate security is via side-channels attacks~\cite{aviram2010determinating}. 
These exploit unintended information leakage due to the sharing of physical resources (such as CPU duty cycles, power consumption, memory allocation).
For example, a malicious software in one VM can try to understand patterns in memory allocation of another co-hosted VM without the need of compromising the hypervisor. 
One of the first real examples of such attacks has been shown in~\cite{zhang2012cross}, where the researchers demonstrated how to extract private keys from an adjacent VM. 

Finally, deleted data in one VM can be resurrected from another VM sharing the same storage device in a Data Scavenging attack~\cite{hashizume2013analysis}) or the whole cloud infrastructure can be mapped to locate a particular target VM to be attacked later~\cite{ristenpart2009hey}. 
In addition, the volatile nature of the cloud resources makes it difficult to detect and investigate attacks.  When VMs are turned off, their resources (CPU, RAM, storage, etc.) become available to other VMs in the cloud~\cite{hay2011storm}, making it difficult to track processes and gather forensic evidence.

\begin{recommendation}
While we believe that the cloud is the future of massive data collection studies, the current situation still presents some technical difficulties that need to be addressed. 
For this reason, in section~\ref{sec:dataControl} we recommend focus on methods to control data treatment (\textit{information flow} and \textit{data expiration}) for remote storage systems in order to ensure user compliance with privacy agreements. 
For the time being, data security can  be improved by combining some of the solutions previously outlined. 
For example, information should be stored encrypted when still on the smartphones and then transported to the servers via secure channels (HTTPS). 
After processing on researcher machines, they can be re-encrypted prior to cloud storage.
\end{recommendation}

\section{Privacy and Datasets}\label{sec:datasets}
The datasets created in massive data collection studies often contain very sensitive information about the participants. 
A common definition of Personally Identifiable Information (PII) is `any information about an individual maintained by an agency, including any information that can be used to distinguish or trace an individual's identity, such as name, social security number, date and place of birth, mother's maiden name, or biometric records; and  any other information that is linked or linkable to an individual, such as medical, educational, financial, and employment information\footnote{NIST Special Publication 800-122 \url{http://csrc.nist.gov/publications/nistpubs/800-122/sp800-122.pdf}}. 
It is researchers' responsibility to protect users' PIIs, and consequently their privacy when disclosing the data to public scrutiny~\cite{narayanan2008robust,barbaro2006face,sweeney2000simple} and to guarantee that the services provided will not be malicious~\cite{ypodimatopoulos2010follow,popa2009vpriv,hull2006cartel}. 
PIIs can be removed, hidden in group statistics, or modified to become less obvious and recognizable to others, but the definition of PII is context-dependent, sometimes making it very difficult to select which information needs to be purged. 
In addition, modern algorithms can re-identify individuals, even if no apparent PII is published~\cite{raij2011privacy,altshuler2011stealing,lane2012feasibility,de2013predicting,de2013unique}. Attempts to provide anonymity often decrease the data utility by reducing resolution or introducing noise, resulting in  privacy-utility trade-off~\cite{li2009tradeoff}.

\subsection{Privacy Implementations}\label{sec:privacyImplementations}
When sensitive information is shared with untrusted parties, technical mechanisms can be employed to enhance the privacy of participants by transforming the original data. 
We present two common ways to augment users' privacy: noise and anonymization, and discuss developments in applied homomorphic encryption. 
For a classification of different privacy implementation scenarios---such as multiple, sequential, continuous, or collaborative data publishing---see~\cite{fung2010privacy}.

\subsubsection{Noise}\label{sec:noise}
A difficult trade-off for researchers is how to provide third parties with accurate statistics on the collected data while protecting the privacy of the individuals in the records: how to address the problem of \textit{statistical disclosure control}. 
Although there is a large literature on the topic, the variety of techniques can be coarsely divided into two families: approaches that introduce noise directly in the database (\textit{data perturbation models} or \textit{offline methods}) and a second group, that interactively modifies the database queries (\textit{online methods}). 

Early examples of the privacy-aware data-mining aggregations can be found in~\cite{agrawal2000privacy}, where the authors considered building a decision-tree classifier from training data with perturbed values of the individual records, and showed that it is possible to estimate the distribution of the original data values. 
This implies that it is possible to build classifiers whose accuracy is comparable to the accuracy of classifiers trained on the original data. 
In~\cite{agrawal2001design} the authors showed an Expectation Maximization (EM) algorithm for distribution reconstruction, providing robust estimates of the original distribution given that a large amount of data is available. 
A different approach was taken in~\cite{evfimievski2003limiting}, where the authors presented a new formulation of privacy breaches and proposed a methodology for limiting them. 
The method, dubbed \emph{amplification}, makes it possible to guarantee limits on privacy breaches without any knowledge of the distribution of the original data. 
An interesting work on the trade-off between privacy and usability of the perturbed (noisy) statistical databases has been included in~\cite{dinur2003revealing}.
These results were extended in~\cite{dwork2004privacy}, where the authors investigated the possibility of a sublinear number of queries on the database which would guarantee privacy, and thuse extend the framework. 
A second work consolidated discoveries from~\cite{dinur2003revealing}, demonstrating the possibility to create a statistical database in which a trusted administrator introduces noise to the query responses with the goal of maintaining privacy of individual database entries. 
In~\cite{blum2005practical} the authors showed that this can be achieved using a surprisingly small amount of noise --- much less than the sampling error --- as long as the total number of queries is sublinear in the number of database rows. 

A different approach was evaluated by Dwork et al. in~\cite{dwork2006our}, where an efficient distributed protocol for generating shares of random noise and securing against malicious participants was described.
The innovation of this method was the distributed implementation of the privacy-preserving statistical database with noise generation. 
In these databases, privacy is obtained by perturbing the true answer to a database query by the addition of a small amount of Gaussian or exponentially distributed random noise, effectively eliminating the need for a trusted database administrator. 
Finally, in~\cite{chawla2005toward} Chawla and Dwork proposed a definition of privacy (and privacy compromise) for statistical databases, together with a method for describing and comparing the privacy offered by specific sanitization techniques. 
They obtained several privacy results using two different sanitization techniques, and then showed how to combine them via cross training. 
This work was advanced in a more recent study~\cite{chawla2012privacy}, where the scope of the techniques was extended to a broad class of distributions and randomization by a histogram construction to preserve spatial characteristics of the data, allowing approximation of various quantities of interest in a privacy-preserving fashion, e.g. cost of the minimum spanning tree on the data.  

\subsubsection{Anonymization}\label{sec:anonymization}
The most common practice in the data anonymization field is to one-way hash all the PIIs such as names, MAC addresses, phone numbers, etc. 
This breaks the direct link between the user in the given dataset and other, possibly public datasets (e.g. Facebook profile). 
The raw data can be uploaded from the smartphone to an intermediate proxy server, where the hashing is performed.
This technique is used in the \textit{LDCC} study.
Once anonymized, the data can be transferred to a second server to which researchers have access. 
A more privacy-preserving option is to hash the data directly on the smartphones and then upload the results onto the server for analysis; a solution used in many of the MIT studies~\cite{aharony2011social,madan2011pervasive,madan2010social,madan2011sensing}
and in the \textit{Copenhagen Networks Study}~\cite{stopczynski2014measuring}. 
In principle, hashing does not reduce the quality of the data (provided that it is consistent within the dataset), but it makes it easier to control which data are collected about the user and where they come from.  
It may, however, reduce the utility if the labels themselves are to be analyzed (e.g. mining the area codes of phone numbers).
In this case the mining should happen before hashing, and hashes should be stored together with low-dimensional results computed from the raw data. 
Some types of raw data -- such as audio samples -- can be \emph{obfuscated} directly on the phone without losing the usability before being uploaded~\cite{kiukkonen2010towards,olguin2009sensible}.
Hashing PIIs does not guarantee that users cannot be identified in the dataset~\cite{barbaro2006face,sweeney2000simple,narayanan2008robust}.
Section \ref{sec:attacksPrivacy} contains examples of methods that revert and break such `anonymity' through the use of auxiliary (non-hashed) information, de facto revealing individuals' identities.

A frequent method employed for anonymization is ensuring $k$-anonymity~\cite{sweeney2002k} for a published database. 
This technique ensures it is not possible to distinguish a particular user from at least $k-1$ people in the same dataset. 
\textit{AnonySense} and the platform developed for the \textit{LDCC} both create $k$-anonymous different-sized tiles to preserve users' location privacy, outputting a geographic region containing at least $k-1$ people instead of a single user's location. 
Nevertheless, later studies have shown how this property is not well suited for a privacy metric~\cite{shokri2011quantifying}. Machanavajjhala et al. tried to solve the $k$-anonymity weaknesses with a different privacy notion called $l$-diversity~\cite{machanavajjhala2007diversity}. 
This was extended by Li et al. by proposing a third metric, $t$-closeness~\cite{li2007t}. 

While today's anonymization techniques might be considered robust enough in providing privacy to the users~\cite{cavoukian2011dispelling}, our survey contains methods that manage to re-identify participants in anonymized datasets (see section~\ref{sec:attacksPrivacy}).

\subsubsection{Homomorphic Encryption}\label{sec:homomorphic}
Homomorphic encryption is a cryptographic technique enabling computation with encrypted data: operations in the encrypted domain correspond to meaningful operations in the plaintext domain~\cite{rivest1978data,gentry2009fully}. 
This means users can allow other parties to perform operations on their encrypted data without exposing the original plaintext, thereby limiting the sensitive data leakage.

This technique can find application in health-related studies, where patients' data should remain anonymous before, during, and after the studies, with only authorized personnel having access to the clinical data. 
\textit{Data holders} (hospitals) can send encrypted information on behalf of \textit{data producers} (patient) to untrusted entities (e.g. researchers and insurance companies), who process them without revealing the data content, as formalized by \textit{mHealth}, an early conceptual framework. 
\textit{HICCUPS} is a concrete prototype that allows researchers to submit medical requests to a query aggregator that routes them to the respective caregivers. 
The caregivers compute the requested operations using sensitive patients' data and send the reply to the aggregator in encrypted form. 
The aggregator then combines all the answers and delivers the aggregate statistics to the researchers. 
A different use of homomorphic encryption to preserve users' privacy is demonstrated by \textit{VPriv}. 
In this framework the central server first collects anonymous tickets produced when cars exit the highways, then, by homomorphic transformations, it computes the total amount that each driver has to pay at the end of the month. 

Secure two-party computation can be achieved with homomorphic encryption when both parties want to protect their secrets during the computations: none of the involved entities needs to disclose its own data to the other, at the same time they achieve the desired result. 
In~\cite{franz2012secure} the researchers applied this technique to private genome analysis. 
A health care provider holds a patient's secret genomic data, while a bioengineering company has a secret software that can identify possible mutations. 
Both want to achieve a common goal (analyze the genes and identify the correct treatment) without revealing their respective secrets: the health care provider is not allowed to disclose patient's genomic data; the company wants to keep formulas secret for business reasons. 

Recently, much effort has been made in building more efficient homomorphic cryptosystems (e.g.~\cite{tebaa2012homomorphic,naehrig2011can}). 
It is still hard to foresee whether or when the results will be practical for massive data collection frameworks.

\subsection{Attacks against Privacy}\label{sec:attacksPrivacy}
While researchers mine the data for scientific reasons, malicious users can misuse it in order to perpetrate `illegitimate acquisition and analysis of people's information'~\cite{altshuler2011stealing}.

Like scientists, reality thieves aim to decode human behavior such as everyday life patterns~\cite{shokri2011quantifying},~friendship relations \cite{quercia2009friendsensing,eagle2009inferring}, political opinions~\cite{madan2011pervasive}, or purchasing profiles\footnote{\url{http://adage.com/article/digital/facebook-partner-acxiom-epsilon-match-store-purchases-user-profiles/239967/}}.
Businesses invest in data mining algorithms in order to make high-quality predictions of customer purchases, while others are interested in analyzing competitors' customer profiles~\cite{valueOfPrivacy}. 

Privacy scandals such as \textit{Netflix Prize}, \textit{AOL searcher}~\cite{barbaro2006face}, and the \textit{Governor’s of Massachusetts health records}~\cite{sweeney2000simple} show that the de-identification of data elements is often insufficient, as it may be reversed, revealing the original individuals' identities. 

A common approach to data re-identification is to compare `anonymized' datasets against the publicly available ones; such schemes are called \textit{side-channel information} or \textit{auxiliary data} attacks.
For such attacks, online social networks are excellent sources of auxiliary data~\cite{lane2012feasibility}. 
In recent studies~\cite{srivatsa2012deanonymizing,mislove2010you}, researchers have shown that users in anonymized datasets may be re-identified studying their interpersonal connections  on public websites such as Facebook, LinkedIn, Twitter, Foursquare, and others. 
Researchers identified similar patterns connecting a pseudonym in the anonymized dataset to the user's real identity in a public dataset. 
It is difficult to thwart such side-channel attacks, as even a small number of low-resolution datapoints from the dataset can be used to uniquely fingerprint a user as shown in~\cite{de2013unique}, and obtaining a similar small number of datapoints about an individual from public sources may be equally easy. 
Using examples from the studies we have surveyed, users' anonymity might be compromised in \textit{VPriv} and \textit{CarTel} every time only a single car is on a highway, because it is possible to link the anonymous packets reaching the server to that (unique) car. 
The same type of consideration is valid for \textit{AnonySense} and \textit{VirtualWalls}; if only one user is performing the sensing task (first platform) or if only one user is located inside a room at a certain time (second platform). 
\textit{CenceMe}, \textit{FollowMe}, and \textit{Locaccino} allow the users to update their daily habits. 
Studying symmetries (e.g. in commuting) and frequencies (e.g. meals or weekly workouts) of these behaviors, it is possible to discover underlying patterns and perpetrate attacks against users' privacy~\cite{lane2012feasibility}. 
More generally, it is often the case the data released by researches is not the source of privacy issues, but the unexpected inferences that can be drawn from it~\cite{tene2012big}. 

Internal linking~\cite{lane2012feasibility} is another approach that aims to connect different interactions of one user within the same system. 
For example, in~\cite{kapadia2008anonysense} two reports uploaded by a single user might be linked on the basis of their timing. 

Data collected by seemingly innocuous sensors can be exploited to infer physiological and psychological states, addictions (illicit drugs, smoking, drinking), and other private behaviors. 
Time and temperature can be correlated to decrease location privacy~\cite{aggarwalsocial}, while accelerometers and gyroscopes can be used to track geographic position~\cite{raij2011privacy} or even to infer peoples' mood~\cite{lane2012feasibility}.
A recent study showed that accurate estimates of personal attributes (such as IQ levels, political views, substance use, etc.) can be inferred from the Facebook Likes, which are publicly available by default~\cite{kosinski2013private}.

These threats are exacerbated because the general public is often unaware of what can be inferred from seemingly harmless data~\cite{mahato2008implicit} and of smartphone sensing capabilities~\cite{klasnja2009exploring}. 
For example, participants of the \textit{Bluemusic} experiment did not show any concerns related to options such as `recording all day, everyday' and `store indefinitely on their mobile phone' related to data collected by accelerometers because this sensor is perceived as `not particularly sensitive'~\cite{mahato2008implicit}. 

While protecting the privacy of an individual user is difficult, data with a network structure make it even more challenging.
The nodes can be identified on the basis of the network structure and knowledge of some attributes of the neighbors of the node~\cite{zhou2008preserving}.
Even when using conventional de-identification techniques, the information embedded in the network structure can be sufficient to find the identity of the user.
Anonymization of network data to prevent these attacks is difficult, and requires careful changes in the disclosed network structure to reduce the possibility of re-identification while keeping the utility of the network data.
Cheng et al. showed in~\cite{cheng2010k} that k-isomorphism, anonymization based on formation of k pairwise isomorphic subgraphs is both sufficient and necessary to protect against re-identification.
The problem of constructing the subgraphs is, however, NP-hard and such operation can significantly reduce the utility of the network data.

Protecting against re-identification in the network is only part of the solution; perhaps even more important is protection against attribute disclosure.
The latter happens when certain attributes the user would like to keep private are disclosed due to the user's position in the network.
Davis et al. in~\cite{davis2011inferring} recovered the location of untagged tweets using the network structure of the followers and their geo-tagged tweets.
In a dataset related to a particular topic on Twitter, they showed an increase of 45\% in the number of messages that could be located.
Sensitive attribute inference was also showed in~\cite{zheleva2009join}, where the authors mixed private and public profiles in social networks, using publicly available friendship structure and group memberships to recover hidden values of the user, such as gender, religion, and origin.
Using models of variable complexity and network properties such as autocorrelation, caused by homophily or influence, they were able to successfully show significant leakage of information from friendships and groups users join.
Chester and Srivastava introduced in~\cite{chester2011social} an approach to anonymize labeled social networks.
Called $\alpha$-nearness, it produces label distribution in every neighborhood of the graph close (within $\alpha$) to that of the entire network by producing a graph with an augmented set of edges.

Privacy in network data is arguably the hardest challenge of the entire domain.
It touches the legal and ethical dimensions (who owns the story about me), informed consent (how my assent influences others), and data control (even if my data is deleted, it may still be recovered from other users).
Addressing the issue in full is beyond the scope of this article.
We recommend that these problems need to be considered in addition to the privacy of a single user, which is a necessary first step.

The consequences of loss of sensitive information can be long lasting; it is almost impossible to change personal relationships or life patterns\footnote{Although for slightly different reasons, in 2010 Google's Executive Chairman Eric Schmidt suggested automatic changes of virtual identities to  reduce the limitless power of the `database of ruin' \url{http://online.wsj.com/article/SB10001424052748704901104575423294099527212.html} .} after an \textit{identity theft attack} or to avoid other types of criminal activity (stalking) that might occur because of misuses of behavioral datasets~\cite{altshuler2011stealing}.

\begin{recommendation}
We suggest that the potential for misuse and attacks should be mentioned in the informed consent processes.
Finally, many existing platforms build privacy for the users on the hypothesis that techniques like $k$-anonymity and protections against traffic analysis or side channels will possibly be added in future, but at the time of writing, such techniques have not yet been integrated in any study. 
As practitioners of massive data collection, we feel that one of the most important challenges of future studies will be to build common frameworks that provide holistic, reusable solutions addressing privacy concerns.
\end{recommendation}

\section{Information Ownership and Disclosure}\label{sec:informationOwnership}
An attractive solution to some of the issues listed previously would be frameworks that can guarantee the users ways to control at any time who is in possession of their data and for what reason, but past attempts at creating such digital rights management (DRM) systems for privacy purposes did not produce the expected results. 
Further, without trusted computing bases there is no practical way of enforcing that data are not to be retained/copied or forwarded to third parties~\cite{hong2004architecture,qiu1985trusted}. 

One problem remains in the issue of whether the users allow researchers to physically own the data stored on the universities' servers (see section~\ref{sec:centralServer}) or simply allow researchers to use the information stored as personal datasets (section~\ref{sec:cloud} and~\ref{sec:sharing}).
Users must trust researchers and service providers to properly manage their personal information as agreed and not to expose any sensitive data to unauthorized parties. 
In the following sections we provide further details about two main information disclosure scenarios: the explicit and conscious process of distributing information to other parties, and techniques to control data disclosure, such as information flow control and data expiration systems.

\subsection{Sharing}\label{sec:sharing}
Individuals' notions of information sensitivity, and therefore sharing patterns, vary~\cite{ypodimatopoulos2010follow,shabajee2006informed,lindqvist2011m,toch2010empirical}.  Some information, however, should always be perceived as sensitive and requires special attention. 
Examples include health-related, financial, or location data.
The sensitivity of information is often proportional to its research value, making users reluctant to share such data~\cite{kotz2011threat,raij2011privacy,klasnja2009exploring}. 
For example, recent studies have demonstrated that social networks can be mined in order to discover psychological states~\cite{lane2012feasibility,de2013predicting}, which can be later used to detect unusual patterns and prevent or treat psychological disorders (social anxiety, feelings of solitude, etc.).

\subsubsection{Sharing with other users}
Social networks and smartphone applications such as \textit{CenceMe}, \textit{Bluemusic}, \textit{FollowMe}, and \textit{Cityware} show people are comfortable with the idea of sharing personal information with friends~\cite{korth2011interdisciplinary}.
However, in many cases the data sharing options lack granularity. 
For example, users of \textit{CenceMe} and \textit{FollowMe} can unfriend other participants and thus resize the sharing set, while users of \textit{Cityware} or \textit{GreenGPS} need to switch off the Bluetooth/GPS device to avoid being tracked. 
More fine-grained systems exist, in which the users can visualize and edit boundaries and resolution of the collected information before sharing it with others~\cite{hong2004architecture,kapadia2007virtual,kiukkonen2010towards,toch2010empirical}.

Location sharing, in particular, is a multifaceted topic. 
Today, users can instantaneously share their location through an increasing number of services: using native applications (Foursquare, Google+), by means of social network websites (Facebook, Twitter) and within social experiments~\cite{ypodimatopoulos2010follow,miluzzo2008sensing,kostakos2008cityware,toch2010empirical,lindqvist2011m}. 
The attitude of users towards location sharing varies.
It has been established the users' understanding of policies and risks is quite limited and often self-contradictory~\cite{shakimov2011vis,lindqvist2011m,toch2010empirical,duckham2010moving}; although the majority of users seem to be at ease in sharing their check-ins, they assert to be worried about the `Big-Brother effect' when asked directly~\cite{raij2011privacy,felt2012android,egelman2012choice}. 
These contradictions and the natural complexity of different preferences in location sharing policies raise challenges for researchers.

\begin{recommendation}
We find that better ways to inform users about possible dangers related to sharing data are needed (see section~\ref{sec:permissionAndConsent} and~\ref{sec:datasets}). 
In addition, we believe new dynamic platforms should be created to enable users to visualize and control their information flows. 
One example would be a platform showing how to discern which areas can report users' location and which cannot~\cite{toch2010empirical}.
\end{recommendation}

\subsubsection{Sharing with researchers}
As discussed above, the most common approach for the studies surveyed here is to collect user-data in centralized servers. 
This means that when the data leaves the smartphone, the user effectively loses control over them and cannot be sure, from the technical perspective, that her privacy will be respected. 
The opposite approach is to let the users own the data so they can control, access, and remove them from databases at any moment, as noted in section~\ref{sec:dataSec}. 
A possible way to achieve this is to upload the data from the mobile devices not to a single server, but to personal data stores (e.g.: personal home computers, cloud-based virtual machines) as shown in the architectures in~\cite{hong2004architecture,gemmell2006mylifebits,shakimov2011vis}. 
It is then possible to deploy distributed algorithms capable of running with inputs coming from these nodes, as illustrated by the \textit{Darwin} framework.

While the advantages of cloud computing for the data collection frameworks are numerous (as discussed in section~\ref{sec:cloud}), this architecture is not immune from privacy concerns for users and technical barriers for scientists. 
The former are worried about the confidentiality of their remote information, the latter need practical ways to collect and perform analysis on the data. 

\subsection{Data Control}\label{sec:dataControl}
Controlling ownership of digital data is difficult. 
Whenever images are uploaded to photo-sharing services, or posts are published on social networks, the user loses direct control over the shared data. 
While legally bound by usage agreements and terms of service, service providers are out of user's direct control. 
The open problems in data disclosure include time retention and information flow tracking. 
Today's frameworks try to solve these issues by legal means, such as declarations asserting that `any personal data relating to the participants will not be forwarded to third parties and will be destroyed at the end of the project' \cite{karikoski2011measuring}. 
In this chapter we discuss the state-of-the-art of the technical means to limit information disclosures and possibilities of integration with research frameworks.

\subsubsection{Information Flow Control}
\textit{Information Flow} (IF) is any transfer of information from one entity to another. 
Not all the \textit{flows} are equally desirable, e.g. a sheet from a CIA top-secret folder should not leak to another file of lower clearance (a public website page, for instance).
There are several ways of protecting against information leaks. 
In computer science, Access Control (AC) is the collection of network and system mechanisms that enforce policies to actively control how the data is accessed, by whom, how, and who is accounted for. 

Some attempts at building decentralized information flow control (DIFC) systems to track information flows in distributed environments are \textit{JIF} extensions (such as \textit{Jif/Split} \cite{zdancewic2002programming} and \textit{CIF} \cite{sfaxi2010information}) or \textit{HiStar} extensions like \textit{DStart} \cite{zeldovich2008securing}, which utilize special entities at the endpoint of each machine to enforce information exchange control. 

An interesting approach taken by \textit{Neon} \cite{zhang2010neon} can control, not only information containers (such as files), but also the data written inside the files. 
It is able to track information flows involved in everyday data manipulations, such as cut and paste from one file to another or file compression. 
In these cases, privacy policies about participants' records stored in datasets cannot be laundered on purpose or by mistake. 
\textit{Neon} applies policies at the \textit{byte-level}, so whenever a file is accessed, the policy propagates across the network, to and from storage, maintaining the binding between the original file and derived data to which the policy is automatically extended.

\textit{Privacy Judge} \cite{konings2011privacyjudge} is a browser add-on for online social networks to control access to personal information published online which uses encryption to restrict who should be able to access it. 
Contents are stored on cloud servers in encrypted form and place-holders are positioned in specific parts in the OSN layouts. 
The plug-in automatically decrypts and embeds the information only if the viewer has been granted access. 
The domains of \textit{Privacy Judge} can be extended by similar tools to limit access and disclosure of personal datasets: the participant could remove one subset of his entries from one dataset, affecting all the studies at once. 

The complementary approach to the above systems is to ensure control of information disclosure \textit{a-posteriori}. 
This means that whoever is in possession of the data, or processing them, can be supervised by the users, and therefore each misuse or unwanted disclosure can be detected. 

Such auditing systems include \textit{SilverLine} \cite{mundada2011silverline}, a tracking system for cloud architectures that aims to improve data isolation and track data leaks. 
It can detect if a dataset collected for one experiment is leaked to another co-resident cloud tenant. 
Therefore, users can directly control where their personal information is stored and who has access to it. 

\textit{CloudFence} \cite{pappas2012cloudfence} is another data flow tracking framework in which users can define allowed data flow paths and audit the data treatment monitoring the propagation of sensitive information. 
This  system -- which monitors data storage and service providers (any entity processing the stored data) -- allows the user to spot deviations from the expected data treatment policies and alert them in the event of privacy breaches, inadvertent leaks, or unauthorized access.

For maintenance reasons (backups, virus scanning, troubleshooting, etc.), cloud administrators often need privileges to execute arbitrary commands on the virtual machines. This creates the possibility to modify policies and disclose sensitive information. 
To solve this inconvenience, \textit{H-one} \cite{ganjali2012auditing} creates special logs to record all information flows from the administrator environment to the virtual machines, thereby allowing the users to audit privileged actions.

Monitoring systems like \textit{Silverline}, \textit{CloudFence}, and \textit{H-one} can be deployed for research frameworks to give the users a high degree of confidence in the management of their remote personal information stored and accessed by cloud systems.

\begin{recommendation}
Unfortunately, these solutions are still not easily deployable since a) many of them require Trusted Computing Bases \cite{krohn2007information,shakimov2011vis, demsky2011cross} (to prove trusted hardware, software, and communications) which are not common at the time of writing; b) some require client extensions that reduce usability and might introduce new flaws \cite{konings2011privacyjudge}; and c) covert channel attacks are not defeated by any IF techniques (e.g. screenshots of sensitive data). 
In addition, enforcing information flow policies also needs to take into account incidental (and intentional) human malpractices that can launder the restrictions.
We note that none of the surveyed frameworks provided information flow controls, and only few of them mentioned auditing schemes. 

While this type of user-protection is less deployable than others, we believe that auditing will have a place in future massive data collection backends. 
We also hold that a paradigm shift in data treatment where users will \textit{own} their personal sensitive information, will make such auditing systems more feasible.
\end{recommendation}

\subsubsection{Data Expiration}
As we have recently seen for Google Street View and Facebook, service providers are very reluctant to get rid of collected data\footnote{\url{https://www.schneier.com/blog/archives/2009/09/file_deletion.html},\\ \url{http://www.dailymail.co.uk/sciencetech/article-2179875/Google-admits-STILL-data-Street-View-cars-stole.html}}.
After user deletion requests, service providers prefer to make the data inaccessible -- hidden from view, behind a disabled profile -- rather than physically purging the data from storage. 
To aggravate this situation, data is often cached or archived in multiple backup copies to ensure system reliability. 
Therefore, from the users' perspective it is difficult to be completely certain that every bit of personal information has been deleted. 
Consequences of unlimited data retention can be potentially catastrophic: if private data is perpetually available, then the threat to user privacy becomes permanent~\cite{castelluccia2011ephpub}. 
A solution to this problem is \textit{retroactive privacy}: meaning that data will remain accessible until -- and no longer than -- a specified time-out.

Here we illustrate some of the most interesting approaches to address the data expiration strategy, narrowing our focus to systems that can be integrated with research frameworks. 
The criteria in the selection are a) user control and b) ease of integration with existing cloud storage systems. 
Self-destructing data systems prefer to alter data availability instead of its existence, securing data expiration by making the information unreadable after some time. 
The concepts `self-destructing data' and `assured data deletion' were first presented in~\cite{boneh1996revocable}. 
Data is encrypted with a secret key and stored somewhere to be accessible to authorized entities. 
Then, after the specified time has passed, the corresponding decryption key is deleted, making it impossible to obtain meaningful data back. 
This is a trusted-user approach and relies on the assumption that users do not leak the information through side channels, e.g.~copying protected data into a new non-expiring file. 
Therefore, these systems are not meant to provide protection against disclosure during data lifetime (before expiration), as DRM systems are designed to achieve\footnote{DRMs assume user untrustworthiness limiting the use and/or disclosure of a digital content in all its possible forms e.g.: duplication and sharing.}. 
Self-expiring data systems can be integrated in research frameworks to enhance privacy in sharing data, permitting the participants to create personal-expiring data to share with researchers for only a pre-defined period of time.

Key management---which is the main concern in such systems---can be realized either as a centralized \textit{trusted} entity holding the keys for all the users, or keys can be stored across different nodes in a distributed network where no trusted entity is employed. 

\textit{Ephemerizer}~\cite{perlman2005ephemerizer,perlman2005file} extends the principles outlined in~\cite{boneh1996revocable} to interconnected computers implementing a central server to store the keys with respective time-outs. 
The server periodically checks the keys for their time-out and delivers them only if their time has not yet expired. 
An approach that avoids the necessity for a trusted party is \textit{Vanish}~\cite{geambasu2009vanish}, a distributed solution that spreads the decryption key bits among different hosts: after the usual encryption phase (key creation, file encryption and storing), the key is split into secret shares and distributed across random hosts in a large distributed hash table (DHT) (DHT are decentralized systems that store $<key,value>$ mappings among different nodes in a distributed network). 
The key tells which node is holding the corresponding value/piece of data, allowing value retrieval, given a key. 
According to the secret sharing method~\cite{naor1995visual}, the recovery of $k$ (threshold) shares on $n$ total shares permits the reconstruction of the original key and therefore the decryption. 
What makes the data expire/vanish is the natural turnover (churn) of DHTs (e.g.: Vuze) where nodes are continuously leaving the network making the pieces of a split key disappear after a certain time.  
When there are not enough key shares available in the network, the encrypted data and all its copies become permanently unreadable. 
There are two main limits on the Vanish system.
First, the requirement for a plug-in that manages the keys reduces its usability. 
Secondly, the time resolution for expiration is limited to the natural churn rate of the underlying DHT and it is expensive to extend due to re-encryption and key distribution. 

As pointed out in~\cite{wolchok2010defeating}, the idea of turning the nodes' instability into a vantage point for data expiration might introduce serious problems. 
Attacker can crawl the network and harvest as many stored values as possible from the online nodes before they leave the network. 
Once enough raw material has been collected, the attack can rebuild the decryption key, resuscitating the data.

Based on the same cache-ageing model of Vanish, but immune to that attack, is \textit{EphPub}~\cite{castelluccia2011ephpub}, in which the key distribution mechanism relies on the Domain Name System (DNS) caching mechanism. 
The key bits are distributed to random DNS resolvers on the Internet, which maintain the information in their caches for the specified \textit{Time To Live}. 
This solution is transparent to users and applications, not involving additional infrastructure (a DHT or trusted servers) nor extra software (DHT client).

Another solution for data expiration is \textit{FADE}~\cite{tang2012secure}, a policy access control system that permits users to specify read/write permissions of authorized users and applications other than data lifetime. 
The underlying idea is to decouple the encrypted data from the keys: information can be stored in untrusted cloud storage providers and a \textit{quorum of distributed key managers} guarantees distributed access control permissions for the established period of time.

Given data redundancy and dispersion, it is almost impossible to ensure full control over distributed data, especially when users are directly involved, as it cannot be prevented that human users manually write down the information, memorize it, or take a picture of the screen and share it in a non-secure manner~\cite{konings2011privacyjudge}. 
While everlasting data is generally dangerous in any context, the problem becomes even more important for studies where the amount and the sensitivity of the collected data can be substantial. 
On the other hand, we can expect that even high-resolution data of a certain kind, for example mobility traces, will become less sensitive with the passage of the time. 
As a society, we will need to understand how to deal with massive amounts of high-resolution data existing after their owners have passed away.

\begin{recommendation}
The systems described above can be part of privacy-aware research frameworks that can automatically take care of purging old information from the database. 
Providing users with ways to control sharing schemes and information lifetime might attract more participants, who may be currently be reluctant to share their personal data. 
We would like to emphasize that the mentioned solutions do not provide complete data protection and have been inspected by the scientific community for only a brief period of time.
It is not current practice in the examined frameworks to include the data retention procedures and lifetimes in the user agreements or informed consent. 
While it is still uncertain whether assured deletion and data expiration are technically secure, we note that there are limits beyond which only legal means can guarantee the users the conformity to certain procedures in data management and retention.
The data deletion schemes must, however, allow for studies' replicability; the minimal set of data or preferably features required to replicate published results must be retained. 
\end{recommendation}

\subsubsection{Watermarking}

Digital watermarking is a process of embedding a marker in the data for the purposes of content tracking, authentication, or owner identification, among others.
Embedding the marker aims to preserve the utility of the data for a particular purpose, assuming a certain level of fault-tolerance in the data in a given context. 
The general requirements of watermarking include that it should be 1) imperceptible, not introducing any perceptible artifacts into the data 2) robust, being immune to certain classes of transformations, 3) blind, in that the detections should neither require knowledge of the watermark or the original database, and 4) incrementally updatable~\cite{agrawal2003watermarking}. 

Transactional watermarking, also known as fingerprinting, is a technique of placing different marks in different copies of the data.
This allows the data owner to trace copies of the data and identify the leaks~\cite{cox2000watermarking}.
The attackers are, in our context, researchers and other entities who have either been granted access to the data, but share or use them in an unauthorized way, or have obtained the data from unauthorized sources.
In active attacks, the entity attempts to remove the watermark or to make it undetectable~\cite{cox1998some}.
A special case of active attacks is a collusion attack, in which the attacker combines several copies of the data, each with a different watermark, to construct a watermark-free copy.
Resistance to these attacks is critical for fingerprinting, if there is a possibility that the attacker can gain access to multiple copies of the same data.
This can happen when one entity can request multiple copies of the data, each with a different embedded watermark, or when several entities can collaborate with each other, sharing their datasets.
Whether it is a concern, depends on the data access framework, how the data is accessed and by whom.
If the access is tightly controlled and audited, or the watermark is consistent per entity accessing the data, collusion attack may not be feasible for a single entity.
The parties accessing the data may also be unlikely to conspire to produce a watermark-free copy, for example respected researchers or large companies.
However, if data access is possible, for example, for anonymous users on the Internet, they may be interested in working together to attack the fingerprint.

Various generic techniques of watermarking relational databases have been developed.
Agrawal et al. showed in~\cite{agrawal2003watermarking} how a relational database can be watermarked using certain bit positions of some of the attributes of the tuples.
These locations and values are determined on the basis of a secret key, chosen by the owner of the data, allowing detection of the watermark with high probability.
With enough redundancy in the watermark, it is robust with respect to common database operations, including insertions, updates, and deletions.
Guo et al. improved algorithm to watermark numeric attributes in relational databases in~\cite{guo2006improved}.
They used groups of tuples representing each bit of the watermark, and by being able to identify which group the tuples belong to, achieved a totally blind system, where only the length of the watermark is needed to recover the watermark. 
An example of using fingerprints in large datasets is the American Dataset Program (ADT), designed to produce decoy records for the purpose of fingerprinting identity datasets~\cite{white2006using}.
These decoy entries, created to look realistic and not bias the statistics calculated on the data, can be generated uniquely for every copy of the dataset released.
The records are highly unique and are based on a key chosen by the owner of the data, thus they can always be identified.

Watermarking can be a suitable technique for data control in massive data collection efforts.
Depending on the nature of the data and expected misuses, the effectiveness may vary, however.
We can expect that the problem is often not leaking of the entire dataset, but rather using the data in an unauthorized way.
In such a scenario, when the actual data is never leaked, but covertly used against the users, identifying the source of the data leak may be hard, even with the watermark embedded.
Some interesting solutions can be envisioned in such cases, for example, if the danger is about abusing collected email addresses of the participants for the purpose of spamming, one can embed the decoy records as described in~\cite{white2006using} and monitor the generated email addresses for spam.
As long as the watermark does not significantly reduce the utility of the data, embedding it is a good idea. 
The usefulness may, however, vary in the massive data context.

\subsubsection{Contract Governance}

All technical means employed for ensuring control of data sharing have their limitations.
No system is perfectly secure, covert attacks can always be found, and trying to stop these attacks with only technical means is not feasible; users can be identified, watermarks removed, and data shared beyond controlled systems.
In the end, it is the credible threat of legal and business consequences that has to assist in discouraging widespread abuse.

Data access and sharing can be governed by contract agreement, specifying what can and cannot be done with the data, as well as consequences of the abuses.
It is common practice for researchers to work under agreements that define allowed uses of the massive data~\cite{karikoski2011measuring, kiukkonen2010towards}.
In the Data For Development (D4D) challenge, researchers accessed the data published by telecommunication operator Orange for the purpose of generating novel insights for the public good.
Beyond the technical means of de-identification of the dataset~\cite{sharadanonymizing}, researchers agreed to terms and conditions, stating that they would not attempt the re-identification of users, share the data with third parties, or otherwise abuse the access~\cite{d4d2012anonym}.

Legal means are routinely employed when granting researchers access to data collected by governmental institutions in registers, such as economic data, personal health records, or education records.
In the Netherlands, a contract is signed by the researcher working with the data, while the statement of secrecy is signed by both the researchers and the employing institution~\cite{hundepool2005onsite}.
In Denmark, access to registers, where data from the population are grouped and records are identified by civil registration number, is regulated by the Act on Processing of Personal Data which states that personal data applied for statistical purposes may be disclosed and reused with the permission of the Data Protection Agency~\cite{borchsenius2005new}.
Under this Act, a public authority may impose a duty of non-disclosure on the researchers and even de-identified data must be treated as confidential.
Breaches are punishable by detention or imprisonment.
Similar measures are implemented in Sweden and other Nordic countries~\cite{hjelm2005mona}, with some minor differences in what data exactly can be accessed; for example, in Finland it is not generally possible to access data with personally identifiable information.
The guiding principle in all the countries is that the access to data for research purposes must not put the subjects in danger, including the possibility of re-identification.
Individuals are entitled to be protected from intrusion, at the same time balance must be found with the legitimate needs of society to access data for the public good.

Legal means are sensitive to geography.
For example in Denmark only Danish research environments can be granted access to register data, as it not feasible to effectively enforce contracts abroad~\cite{borchsenius2005new}.
Except for the legal means, no scrambling, grouping, or other statistical randomization techniques are applied to these data.
When accessing register data, it is often a practice to require researchers to physically work on the premises of the managing agency, use designated servers for computation, or use authorized and secure connection~\cite{hundepool2005onsite,borchsenius2005new,hjelm2005mona}.
The data flows are tightly controlled in this context, all performed computations and transfers are logged and audited, and the policy to which researchers agreed is enforced.

In Europe, Council Regulation (EC) No 322/97 of 17 February 1997 on Community Statistics states that the data used for the production of statistics should be considered confidential when they allow for identification of the statistical units, disclosing individual information either directly or indirectly~\cite{regulation31997R0322}.
All the means that can be reasonably used by a third party for such identification have to be considered when deciding whether a statistical unit is identifiable.
This goes beyond protection of PIIs that directly identify users, including any combination of data that makes identification feasible.
Directive 95/46/EC of the European Parliament and of the Council of 24 October 1995 mentions the protection of individuals with regard to the processing of personal data and the free movement of such data (the Data Protection Directive)~\cite{directive199595}.
The Directive applies to electronic personal data and structured manual files, and it talks about processing, which includes collection, storing, merging, changing, sharing, deleting, etc.
The Directive covers fundamental requirements for data processing, including fairness and lawfulnesses, explicit purpose of collection and analysis compatible with the purpose, and keeping the data identifiable for only as long as it is required.
The adoption of the General Data Protection Regulation, superseding the Data Protection Directive and addressing issues such as  cloud computing and data generated by online social networks, is expected for 2014, with a following 2-year transition period (\url{http://ec.europa.eu/justice/data-protection/index_en.htm}).

In Australia, controls on research data are most commonly imposed by government, as a very high proportion of funding is provided by the Commonwealth, State, and Territory governments~\cite{fitzgerald2007building}.
All the aforementioned principles for treatment research data apply in Australia, including requirements for informed consent, removal of identifiable information, and usage only for purposes compatible with the original purpose of data collection.
We mention Australia, because Fitzgerald et al. note in~\cite{fitzgerald2007building} how some research expressed frustration with the limitations imposed by the existing legislation.
Quoting Dr. Richie Gun of the Department of Public Health, University of Adelaide:

\begin{quote}
\emph{In Australia we are now in a uniquely advantageous position to carry out such [cancer] research, as we have mandatory registration of cancers in every State and Territory. 
We therefore have almost complete enumeration of all invasive cancers occurring in Australia, with the potential to carry out epidemiological studies on cancer incidence equal to or better than anywhere else in the world.
Unfortunately privacy laws are impeding access to cancer registry data, so that it is becoming increasingly hard to carry out the linkage of cancer registrations with exposure data.}

\emph{Rulings such as this suggest that we researchers are not to be trusted to protect privacy; that names will be released to outside parties; or that publications will identify individuals.
This might be justified if there were some evidence that researchers have actually misused such data.
Yet where is such evidence?
The fact that there is no evidence of misuse is easily explained: researchers have nothing to gain by providing information and everything to lose.
I know that if it became known that confidential information had been given out from my research team, it would be the end of my research and my career.}
\end{quote}

This statement shows that legal and business means of ensuring proper data flows can be effective in a research context.
When researchers have everything to lose and little to gain from abusing the data, they could be entrusted with high-quality data, even including personally identifiable information, just based on their agreement to not abuse this access.

The IPUMS-International is a global initiative led by the University of Minnesota Population Center to coordinate access to high-resolution census data to researchers~\cite{ruggles2003ipums}.
The projects within the initiative are only undertaken in countries with an explicit understanding between the official statistical institute and the University of Minnesota~\cite{mccaa2005ipums}.
The agreement includes rights of ownership, rights of use, conditions of access, restrictions of use, the protection of confidentiality, security of data, citation of publications, enforcement of violations, sharing of integrated data, and arbitration procedures for resolving disagreements.
Researchers must apply for a license to gain access to the data, and approval is based on whether the data is appropriate for the proposed project, the academic non-commercial affiliation of the researcher, and acceptance of the conditions of use; around one-third of the applications are denied~\cite{mccaa2005ipums}. 
The researchers accessing the data sign a non-disclosure agreement, agreeing to use the data only for non-commercial purposes, not to attempt to identify the individuals, not to report statistics that would allow such identification, and not to redistribute the data to third parties.
Violations of the agreement are subject to professional censure, loss of employment, civil prosecution, and to sanctions against the researcher's institution.
IPUMS notes that with the supervision of IRBs and corresponding bodies in other countries, the lack of incentives for the researchers to identify the individuals, and with researchers' reputation at stake, abusing data is unlikely~\cite{dale2001proposals} and only few allegations have been ever made in the context of registry data.

No matter whether the data are accessed for research or commercial purposes, perfectly secure systems are virtually impossible to build.
Even in top secret government agencies, there must exist people with sufficient access to extract very sensitive data; it is an operational reality.
Whenever the data is shared, some aspects of ensuring the proper processing must rely on legal and business dimensions.
The effectiveness of these varies, depending on the nature of the data and parties accessing them: respected researchers may have a lot to lose and little to gain from abusing the data, large commercial services may try to do it, even if they risk that their misconduct may be discovered.

\section{Privacy Actionable Items}\label{sec:actionables}
Here we present the executive summary for the practitioners of sensor-driven massive data collection studies involving human participants. 

\paragraph{Regulations.} 
When a new study is provisioned, it must follow the regulations of the proper authorities. 
Special attention should be given to the cases where the data may travel across borders, either as part of the study operation (e.g. data storage) or as a part of research collaboration with foreign institutions. 
The requirements and guidelines may differ significantly between countries. 
Furthermore, if the data collection happens in one country and analysis of the dataset happens in another, the data analysis may not be considered a human subjects study, and thus may not require IRB approval. 
The regulations and guidelines of the country where the study is conducted reflect expectations of the participants regarding their data treatment. 
Researchers need to make sure that these are respected, even when the data flows across boarders. 

\paragraph{Informed Consent.} 
Informed consent is the central point of the participant-researcher relation. 
We strongly encourage the publication of the informed consent procedure for the conducted studies, so the best practices can be built around it. 
As a research community, we should be working towards the implementation of Living Informed Consent, where the users are empowered to better understand and revisit their authorizations and their relation with the studies and services in general. 
This relation should ideally last for as long as the users' data exist in the dataset. 
As new techniques for data analysis are introduced and new insights can be gained from the same data, the participants should be made aware of, and possibly in charge of, the secondary use. 
Additionally, we envision a better alignment of business, legal, and technical dimensions of the informed consent, where the user's act of consenting is not only registered for legal purposes, but through technical means creates the required authorizations, e.g.~ OAuth2 tokens.

We do not expect or want users to spend hours staring into the screens to control the access to their data.
Ensuring such metadata is available for the user is a necessary but not sufficient condition for a successful implementation of Living Informed Consent.
Business, legal, and technical dimensions must be aligned to make the presented information accountable and actionable.
The presentation layer must be carefully designed, with the UI as simple as possible, but not simpler than necessary to convey the privacy message.
Finally, providing access to the metadata about data collection and usage, creates an opportunity for third parties to help the user to analyze, audit, understand, and change the privacy settings.
There is a danger that providing raw access to the data and metadata will create an illusion of user control.
Still, the improvement of the users' privacy has to start somewhere, and even if the initial implementations of Living Informed Consent are not perfect, they will create an opportunity to build better solutions upon them.

\paragraph{Security.} 
Security of the data is crucial for ensuring privacy. 
Moving into the cloud may require close examination of the infrastructure provider's policy, as well as the technical solutions limiting access to the data in the shared-resources environment. 
One of the solutions is to encrypt the data on a server physically owned by the research unit conducting the studies, and only then pushing it into the cloud. 
There are different levels of encryption granularity that may be used. 
The less structure is preserved in the data, the less information can potentially leak, but at the same time, less meaningful queries or data processing can be executed on the encrypted data. 
It may be an option to encrypt (or one-way hash) only PIIs, keeping the structure of the data and certain raw values (e.g. location coordinates).
This allows for effective data querying in the database, but at the same time can expose the information about the participants (the cloud learns where the participants are, but doesn't learn their names or other PIIs). 
If the data is encrypted in its entirety, effective queries become difficult and may not be feasible. 

\paragraph{Privacy Implementation.} 
The data collected is only valuable if it can be analyzed. 
It is often desired or necessary to share the data in some form with third parties, some of whom may be hostile. 
As it has been shown in multiple cases, de-identification, introducing the noise, implementing $k-anonymity$ etc. may be insufficient against an educated and determined attacker (\nameref{sec:attacksPrivacy} section). 
We can expect this problem to grow, as more publicly available information is exposed by the participants in different (non-study-related) channels. 
If it is sufficient to know several locations to uniquely identify the user~\cite{de2013unique}, all the other data gathered in the study can then also be linked to the real person. 
Firstly, we suggest that good practice is to make the type of de-identification performed on the dataset public, helping to establish common practices and understanding of what works and what does not. 
One of the possible directions is to move away from copying the entire datasets for later offline analysis, which is becoming impractical anyway, due to large size and real-time nature of the data. 
Instead, researchers and other services can interact with the data through APIs that allow for control and accountability. 
If data dimensionality is additionally reduced before flowing through the API (e.g. city-level location of the user rather than raw GPS trace), privacy can be managed in a more robust way.

\paragraph{Contract Governance.}

The technical means employed for ensuring proper data processing, have their limitations.
When the data is shared and used, it should be done under contracts defining researchers' and services' obligations and allowed uses.
The threat of legal and business consequences should play a crucial role in implementing participants' privacy.
Every party accessing the data should agree not to try to re-identify the users or otherwise abuse the data.
Although this may seem naïve, such agreements can be the most powerful factor in limiting abuses; in many cases researchers or other third parties have little to gain in abusing the data, while risking their reputation and the chance of lawsuits.
It is desired to watermark the data separately for each share in order to support auditing.
Contractual and watermarking practices have been adopted, to certain extent, in sharing medical, biological, and financial data.
Depending on the nature of the data or pre-existing relation between primary researchers and the third party, the contract may need to be of a certain complexity combined with a practical auditing system.
There is always a balance to be found between the difficulty of re-identification, potential harm to the participants, and gain from the conducted research.
When the overhead of establishing legal contracts with all parties accessing the data is prohibitive, a minimal End-User License Agreement (EULA) should be used, where the researchers promise not to attempt re-identification, abuse the data, or share them further.
Similarly to informed consent forms, we suggest making public the forms signed by parties accessing the data in the studies (contracts and EULAs), in order to establish best practices for this process.

\section*{Conclusions}\label{sec:conclusions}

In this article we have reviewed the privacy-related practices in existing sensor-driven studies of massive data collection involving human participants.  
It is clear that the size and resolution of studies grows quickly and for this reason the collected data should be considered increasingly sensitive.
As the sensitivity of the data grows, better tools to support the privacy of the participants have to be developed.

In massive human data collection, the main concerns are currently technical solutions to scalability and data availability. 
Often, making the data available to researchers takes precedence over creating the right solutions for the many aspects of privacy that need to be addressed.
Here, we have identified a number of concrete domains, which are crucial for implementation of good privacy practices. 
These include informed consent management, data security, auditing, or controlling information flows.
We have reviewed tools that can support some of these requirements and noted that, for various reasons, they are not widely used.
We feel that the adoption of reusable and common systems is necessary to improve the privacy situation in the field. 
Reusable tools and systems, as well as shared standards, will allow scientists and engineers to focus on research questions, rather than re-inventing the complex elements of privacy in a new era of massive data collection.

The underlying driver of the change has to be the end-user expression of informed consent and resulting orchestration of data processing in business, legal, and technical domains.
In Living Informed Consent, participants in long-lasting studies with complex data flows will be able to express their preferences and monitor the results.
Going beyond existing practices of obtaining the consent, the users have to be recognized as key stakeholders in the studies.
In addition to promoting better privacy, this will be beneficial for the research community as well, allowing access to more data from the users, rather than starting every collection from scratch.
A key aspect of improving privacy has to be contract governance.
Every flow, every data access should ideally be accounted for and possible to audit, matching it with the user authorizations existing at that time.
Creating such systems, with perfect cooperation of technical and legal solutions is not easy.
In every case, the threat of de-identification, potential harm to the users, and the reputation of the parties accessing the data should be balanced with the gains from the conducted research.
Just as in any other science, robust, efficient, and workable privacy solutions have to be developed iteratively, as a process. 

Here we laid the groundwork for the privacy management change process that can support such transformation.
Although the article focused on the privacy practices in the research context, many of the identified challenges and solutions apply to other domains where the massive data collection takes place, including business and governmental contexts.
Using the review as a starting point, we have constructed a list the most important challenges to overcome in data collection efforts.
We are hopeful users will be put in charge of the data authorizations, and in the near future of the data itself. 
This will be a fundamental shift in the way science is done, and we can expect many studies will not necessarily involve actual data collection.  
Rather, the studies will become applications of already existing user data, managed and monitored by the users themselves.
This is an important step towards the New Deal on Data, ensuring that the required data is available for the public good~\cite{pentland2009reality}.




\bibliography{bibliography} 



\end{document}